\documentclass[%
 reprint,
superscriptaddress,
nofootinbib,
 amsmath,amssymb,
 aps,
 longbibliography,
]{revtex4-1}

\usepackage{multirow,makecell}
\usepackage[table]{xcolor}
\usepackage{mathptmx}
\usepackage{graphicx}
\usepackage{dcolumn}
\usepackage{bm}
\usepackage[normalem]{ulem}  

\usepackage{hyperref}
\hypersetup{
    colorlinks=true,
    linkcolor=blue,
    filecolor=blue,      
    urlcolor=blue,
    citecolor=blue,
}

\begin{document}

\preprint{APS/123-QED}

\title{Dynamics of a Grain-Scale Intruder in a 2D Granular Medium with and without Basal Friction}

\author{Ryan Kozlowski}
\affiliation{Department of Physics, Duke University, Durham, North Carolina 27708, USA}

\author{C. Manuel Carlevaro}
\affiliation{Instituto de F\'isica de L\'iquidos y Sistemas Biol\'ogicos, CONICET, 59 789, 1900 La Plata, Argentina \\
and Dpto. Ing. Mec\'anica, Universidad Tecnol\'ogica Nacional, Facultad Regional La Plata, Av. 60 Esq. 124, La Plata, 1900, Argentina}

\author{Karen E. Daniels}
\affiliation{%
Department of Physics, North Carolina State University, Raleigh, North Carolina 27695, USA
}%

\author{Lou Kondic}
\affiliation{%
 Department of Mathematical Sciences and Center for Applied Mathematics and Statistics, New Jersey Institute of Technology, Newark, New Jersey 07102, USA
}%

\author{Luis A. Pugnaloni} 
\affiliation{Dpto. de F\'isica, Fac. Ciencias Exactas y Naturales, Universidad Nacional de La Pampa, CONICET, Uruguay 151, 6300 Santa Rosa (La Pampa), Argentina}

\author{Joshua E.~S.~Socolar}
\affiliation{Department of Physics, Duke University, Durham, North Carolina 27708, USA}

\author{Hu Zheng}
\affiliation{Department of Physics, Duke University, Durham, North Carolina 27708, USA}%
\affiliation{Department of Geotechnical Engineering, College of Civil Engineering, Tongji University, Shanghai, 200092, China}

\author{Robert P. Behringer}\thanks{Robert (Bob) Behringer passed away on July 10, 2018.  He made important contributions to the design of the experiments reported here, as well as the formation of the collaborative team.
}

\affiliation{Department of Physics, Duke University, Durham, North Carolina 27708, USA}%

\date{\today}

\begin{abstract}
We report on a series of experiments in which a grain-sized intruder is pushed by a spring through a 2D granular material comprised of photoelastic disks in a Couette geometry.  We study the intruder dynamics as a function of packing fraction for two types of supporting substrates: a frictional glass plate and a layer of water for which basal friction forces are negligible.  We observe two dynamical regimes: intermittent flow, in which the intruder moves freely most of the time but occasionally gets stuck, and stick-slip dynamics, in which the intruder advances via a sequence of distinct, rapid events. When basal friction is present, we observe a smooth crossover between the two regimes as a function of packing fraction, and we find that reducing the interparticle friction coefficient causes the stick-slip regime to shift to higher packing fractions.  When basal friction is eliminated, we observe intermittent flow at all accessible packing fractions.  For all cases, we present results for the statistics of stick events, the intruder velocity, and the force exerted on the intruder by the grains.  Our results indicate the qualitative importance of basal friction at high packing fractions and suggest a possible connection between intruder dynamics in a static material and clogging dynamics in granular flows.

\end{abstract}

\pacs{Valid PACS appear here}

\maketitle

\section{Introduction}

Driven granular media exhibit a variety of dynamical behaviors depending on the loading mechanism and properties of the granular material, such as interparticle friction and packing fraction \cite{granularslgjaeger}. It is common for continuous forcing of the driver to result in periodic fluctuations in velocity \cite{stickslipcracklingagheal}, periodic stick-slip behavior \cite{frictiongranularlayersnasuno,stickslipalbert,granularfaultdanielshayman,SimGranularSeismicPicaCiamarra2011}, irregular stick-slip behavior that has been described in some cases as critical \cite{stickslipcracklingagheal,stickslipcouetteDalton}, or mode-switching between periodic and aperiodic regimes \cite{granularfaultdanielshayman}. These phenomena have been observed in experiments and corresponding simulations in which the driving mechanism spans the system, as in boundary shear \cite{granularfaultdanielshayman,localglobalavalanchesbares,SimGranularSeismicPicaCiamarra2011}, or is significantly larger than a single grain, as in the cases of sliders pulled across the surface \cite{stickslipnasuno,stickslipcracklingagheal,SimGranularSeismicPicaCiamarra2011} or rods inserted into the bulk \cite{stickslipalbert,stickslippulldiskoutMetayer}). A key question is how the dynamical behavior may change when the applied stress arises from a grain-scale intruder.

The response of a granular system to single-grain perturbations is of fundamental interest, as it highlights the connections between scales, from single grain rearrangements to force chain creation and destruction to macroscopic energy dissipation and material failure \cite{sticksliptordesillas,probeintruderreichhardt,slowdraggeng}. Early studies of point loads in granular media focused on force propagation in \textit{static} packings \cite{pointloadexperimentsatmangeng,pointloadsimulationgoldenberg}.  More recent experiments have observed the penetration and motion of a grain-scale intruder driven by a constant force or at constant velocity \cite{probeintruderreichhardt,intrudervibrationgdauchot,dragforcecavityformationintruderkolb,slowdraggeng,rheologyintruderexperimentseguin}, and a grain-scale intruder driven through coupling with a spring has been studied in simulations \cite{sticksliptordesillas}. In all cases, the behavior depends strongly on the packing fraction. Experiments on sheared 2D layers of photoelastic disks show that frictional properties, and basal friction in particular, also play an important role in determining macroscopic properties of the system during compression or shear \cite{jammajmudar,shearjamnobfhu,jambyshearbi}. These frictional properties are likely important factors in determining grain-scale intruder dynamics as well.

This paper reports on experiments that elucidate the roles that interparticle friction, basal friction, and packing fraction play in determining intruder dynamics. A single-grain loading mechanism allows us to study the effects of these parameters without the added complication of an averaging of responses over simultaneous direct interactions with many grains.  The system we study consists of an intruder driven through a \textit{confined} channel of grains. In the frame of reference of the intruder, the system shares features of a granular flow through an aperture. One might therefore expect to observe processes analogous to clogging and intermittent flow, with continuous motion of the intruder being interrupted by occasional stick events \cite{clogto,clogconfigsdurian,zuriguel2005jamming,multipleorificesKunte2014}.  For high packing fractions, however, steric constraints lead to strong coupling of the granular rearrangements behind and in front of the intruder, and the motion is expected to resemble more the stick-slip behavior observed in low-speed slider experiments \cite{stickslipcracklingagheal,SimGranularSeismicPicaCiamarra2011,stickslipnasuno}.

We examine the dynamics of a grain-scale intruder that is driven by a torsion spring through an annular channel filled with a 2D bidisperse granular medium. A spring-driven mechanism applies a force to the intruder that increases linearly with time until the granular material yields and the intruder slips. We utilize a Couette geometry to study steady-state dynamics of the intruder after multiple passes through the medium.  In this way, we obtain quantitative measures of the dynamics as the system explores a statistically stationary ensemble of states. In addition, we use photoelastic imaging \cite{photoelastictechniquedaniels,enlighteningforcechainsBares2019} to qualitatively characterize force chain structures in these states. Our results suggest that the presence of basal friction has a strong effect on the characteristics of the force networks associated with stable stick events.

\begin{figure}
    \centering
    \includegraphics[width=0.95\linewidth]{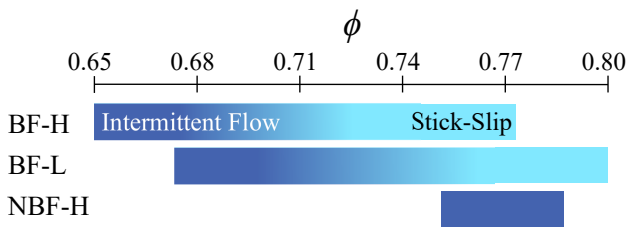}
    \caption{(Color online.) Range of packing fractions $\phi$ explored in each of the three experiments: BF-H: With basal friction and high interparticle friction; BF-L: With basal friction and low interparticle friction; NBF-H: No basal friction and high interparticle friction. Dark represents intermittent flow dynamics and light represents pure stick-slip dynamics.}
    \label{fig:introdiagram}
\end{figure}

We report on three sets of experiments with different coefficients of static friction, $\mu$, for interparticle contacts and different coefficients of static friction, $\mu_{BF}$, for the interaction of a particle with the base. In one case the particles float on water, effectively eliminating the basal friction ($\mu_{BF} = 0$). 
For each set, we vary the packing fraction $\phi$ in a range allowable by the limits of the apparatus (see Sec.~\ref{sec:Methods}). As shown in Fig.~\ref{fig:introdiagram}, when basal friction is present, we find a dynamical regime of intermittent flow at low $\phi$ and aperiodic stick-slip dynamics at high $\phi$, with the crossover from intermittent flow to stick-slip shifting to higher $\phi$ with lower interparticle friction. When basal friction is eliminated, we observe only  intermittent flow, with no stick-slip regime.  These findings experimentally show that basal friction is a key parameter controlling intruder dynamics, and that in a confined geometry clogging-like behavior can occur.

In Sec.~\ref{sec:Methods}, we describe the experimental setup and the parameters used in the three sets of experiments. In Sec.~\ref{sec:results} we discuss the qualitative differences in dynamical behavior and a quantitative measure to distinguish the regimes.

\begin{figure}
    \centering
    \includegraphics[width=0.8\linewidth]{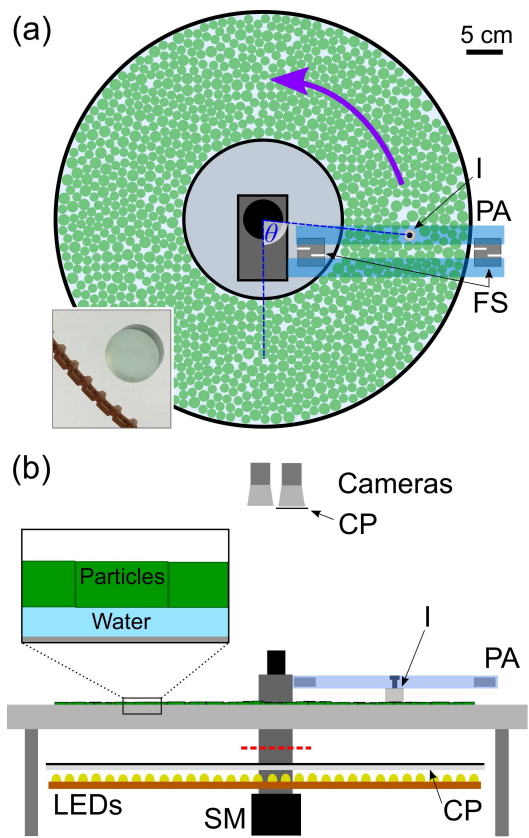}
    \caption{(Color online.) Schematic of apparatus: (a) Top-down (camera) view of apparatus: I = Intruder, PA = Pusher Arm, FS = Force Sensors. The inset shows a closeup image of the ribbed rubber boundary next to a small grain. The angular position $\theta$ of the intruder is measured relative to the start position in a given experimental run. (b) Side view: CP = Circular Polarizer, SM = Stepper Motor. The zoom-in feature shows the approximate height above the table of floating particles. The dashed line (red) just above the CP is the cross-sectional plane with the torque spring, shown in Fig.~\ref{fig:torquespring}.}
    \label{fig:schematic}
\end{figure}

\section{Experiment Design \label{sec:Methods}}

Figure~\ref{fig:schematic} shows a schematic diagram of the apparatus. Approximately 1000 bidisperse polyurethane photoelastic disks are set in an annular channel. The diameters of the particles slightly differ for each set of experiments --- these values are presented in Table~\ref{tab:expparameters}. The number ratio of large to small particles is approximately 1:2.8 in the experiments with high interparticle friction and 1:3.2 in the experiments with low interparticle friction. A bidisperse mixture of particles is used to prevent crystallization \cite{epitomeofdisorderhern}, which we never observe in these experiments. The channel width is 17.8 cm, or approximately 14 small particle diameters. The inner and outer circular boundaries are ribbed rubber with ridges spaced roughly $0.5 \text{ cm}$ apart to reduce particle slip. The particles either sit on the glass table with a friction coefficient $\mu_{BF}$ or float in water above the glass and experience no basal friction forces. In the case with basal friction, the interparticle friction $\mu$ may be high (bare disks) or a low (Teflon\textsuperscript{\textregistered} wrapped disks). Parameter values for each experiment are summarized in Table~\ref{tab:expparameters}.

\setlength\tabcolsep{4pt}
\begin{table*}
\centering
\begin{tabular}{ |c|c|c|c|c| }
 \hline
 Experiment & $\mu_{BF}$ & $\mu$ & $d_s$, $d_l$ ($\pm 0.01$ cm) & Range of $\phi~(\pm 0.008)$ \\ 
 \hline \hline
 (1) Basal friction, high interparticle friction & $0.37 \pm 0.07$ & $1.2 \pm 0.1$ & 1.28, 1.60 & $0.650 - 0.772$ \\
 \hline
 (2) Basal friction, low interparticle friction & $0.34 \pm 0.08$ & $0.18 \pm 0.04$ & 1.34, 1.65 &  $0.677 - 0.796$ \\ 
 \hline
 (3) No basal Friction, high interparticle friction & $0$ & $0.8 \pm 0.1$ & 1.30, 1.62 & $0.750 - 0.786$ \\
 \hline
\end{tabular}
 \caption{Experimental Parameters. Shared parameters for all three experiments are $\kappa_T = 0.431 \pm 0.001 \text{ Nm/rad}$, $\dot{\theta}_{d} = 0.119 \pm 0.006 \text{ rad/s}$, and $d_{int} = 1.59 \pm 0.01 \text{ cm}$. In the sets of experiments with basal friction, $\phi$ was changed in increments of $\sim 0.01$. In the case without basal friction, $\phi$ was usually changed in increments of $\sim 0.002$ except for the three lowest $\phi$, which are separated by increments of $\sim 0.01$. (See, for example, Fig.~\ref{fig:statisticsplots}(a) for the exact packing fractions studied.)}
 \label{tab:expparameters}
\end{table*}

The intruder, a Teflon\textsuperscript{\textregistered} rod of diameter $d_{int} = 1.59 \pm 0.01 \text{ cm}$, which is about the diameter of a large particle in the experiments with basal friction and high interparticle friction, is elevated above the table and therefore has \textit{no basal friction} in all of the sets of experiments. It is rigidly fixed to a cantilever that is attached to a post at the center of the annulus. The intruder's radial distance from the annulus center is fixed at $R = 19.7 \pm 0.1 \text{ cm}$. 

\begin{figure}
    \centering
    \includegraphics[width=0.95\linewidth]{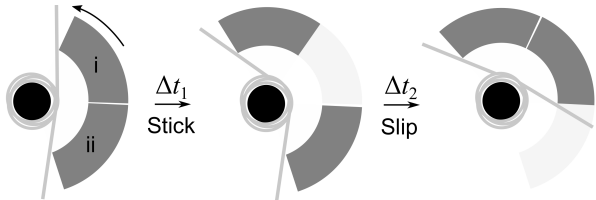}
    \caption{Cross-sectional schematic view of Fig.~\ref{fig:schematic}(b) in the plane marked by the red dotted line: \textbf{i} is rotated at constant angular velocity by the stepper motor and \textbf{ii} is rigidly fixed to the pusher arm. The timescale of the spring's relaxation ($\Delta t_2 \sim 0.4\text{ s}$) is shorter than the timescale of typical stick events ($\Delta t_1$) when pure stick-slip dynamics are observed (light blue in Fig.~\ref{fig:introdiagram}).}
    \label{fig:torquespring}
\end{figure}

\begin{figure}
    \centering
    \includegraphics[width=1\linewidth]{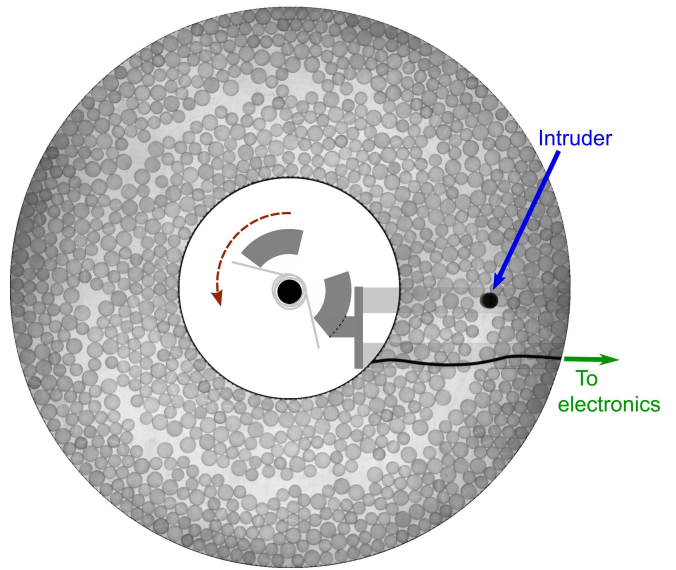}
    \caption{Sample snapshot of an experiment. The intruder is being driven counterclockwise. The black curve is a cable above the particles. The torque spring schematic from Fig.~\ref{fig:torquespring} is shown in the center.}
    \label{fig:sampleimages}
\end{figure}

The cantilever supporting the intruder is fitted with s-beam load cells (JINNUO, JLBS-M2) that measure the force of the granular medium on the intruder at intervals of $0.01$ seconds. A stepper motor (STEPPERONLINE, 24HS34-3008D) and torque spring (stiffness $\kappa_{T} = 0.431 \pm 0.001 \text{ Nm/rad}$) are used to drive the rotation of the cantilever, as shown in Figs.~\ref{fig:torquespring} and~\ref{fig:sampleimages}.  The motor drives one end of the spring at constant angular velocity of $\dot{\theta}_{d} = 0.119 \pm 0.006 \text{ rad/s}$; we define the drive angular velocity direction to be positive.  The other end of the spring pushes the cantilever, driving the intruder through the granular medium. In Fig.~\ref{fig:torquespring}, the wedge labeled \textbf{i} is rotated at constant angular velocity by the stepper motor, and the wedge labeled \textbf{ii} is rigidly fixed to the cantilever.  In a typical stick-slip cycle, the intruder is held in a nominally fixed position by the granular material.  As the motor turns, wedge \textbf{i} advances at a constant rate, building stress in the spring.  When the force on wedge \textbf{ii} reaches the point where the force exerted by the intruder exceeds the yield stress of the given jammed configuration of grains, the intruder, cantilever, and wedge \textbf{ii} slip forward rapidly.   A timescale related to such slip events can be defined as the time required for the compressed spring to fully decompress in the absence of any granular obstacles.  This \textit{inertial timescale} was measured to be $0.40 \pm 0.01\text{ s}$ \footnote{The inertial timescale can be approximated by estimating the moment of inertia of the cantilever as $ML^{2}/3$ and treating the system as a simple harmonic oscillator. In this experiment, $M \approx 0.44 \text{ kg}$ and $L \approx 0.36 \text{ m}$, giving an oscillation period of $1.40 \text{ s}$. The decompression time is thus predicted to be $0.35 \text{ s}$, roughly in agreement with measurements.}.

\begin{figure*}
    \centering
    \includegraphics[width=0.7\linewidth]{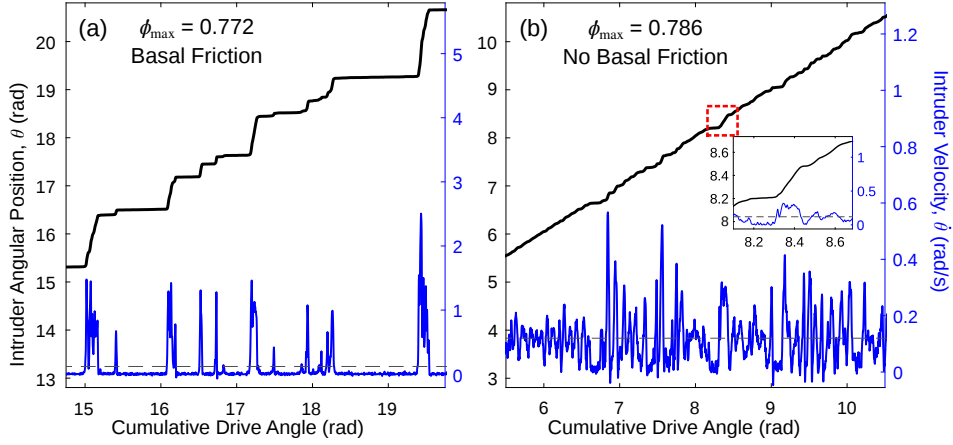}
    \caption{(Color online.) Sample time series of the intruder's angular position $\theta$ (upper curve, black) and intruder velocity $\dot{\theta}$ (lower curve, blue) as a function of the cumulative drive angle of the stepper motor. The dotted line represents the drive velocity $\dot{\theta}_{d}$. (a) Sample from the maximum packing fraction $\phi$ explored with basal friction. (b) Sample from the maximum $\phi$ explored without basal friction. Inset: Zoom in of time series within dotted (red) box.}
    \label{fig:sampletimeseriesplots}
\end{figure*}

The granular medium is illuminated from below by a white light LED panel, with the light first passing through a diffuser and circular polarizer, then through the glass and granular system.  Images are recorded by two cameras above the system.  One camera records the light with no further filtering and is used to measure the particle and intruder locations, which are identified using a circular Hough transform from \textsc{Matlab}\textsuperscript{\textregistered}. The other camera records light that passes through a circular polarizer of opposite handedness to the circular polarizer below the set-up, forming a \textit{dark-field polariscope} \cite{photoelastictechniquedaniels,enlighteningforcechainsBares2019}. This camera is used to estimate the stress in each grain and visualize force networks, both during stick events and rapid slip events. Figure~\ref{fig:sampleimages} displays a sample image from an experiment acquired by the camera without the polarizer. The intruder is tracked throughout experimental runs in intervals of 0.02 seconds. The cumulative angle $\theta$ of the intruder relative to its initial position in an experimental run is recorded. The intruder's angular velocity $\dot{\theta}$ is then computed by a finite difference of the $\theta(t)$ time series. Sample time series of $\theta$ and the corresponding $\dot{\theta}$ are shown in Fig.~\ref{fig:sampletimeseriesplots} as a function of the cumulative drive angle of the stepper motor, which advances with time at a constant rate $\dot{\theta}_{d} \approx 0.12\,$rad/s.

\begin{figure*}
    \centering
    \includegraphics[width=1\linewidth]{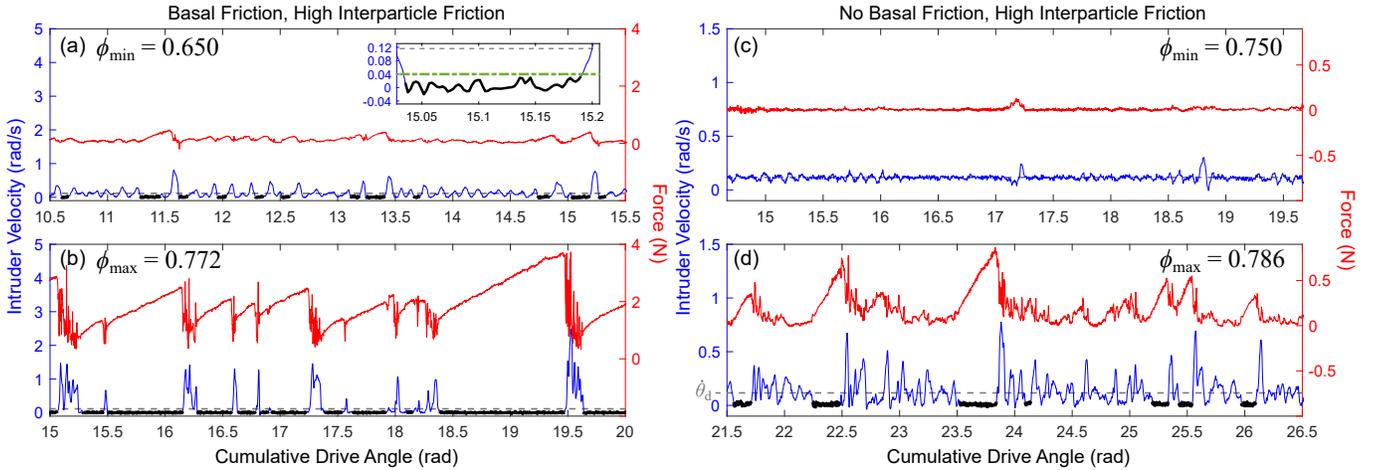}
    \caption{(Color online.) Sample time series for intruder velocity (lower curve, blue) and force (upper curve, red) at the lowest ($\phi_{\text{min}}$) (a,c) and highest  ($\phi_{\text{max}}$) (b,d) packing fractions from the experiments with and without basal friction with high interparticle friction. Thick black segments are detected stick events and the gray dashed line is the drive velocity $\dot{\theta}_{d}$. Inset in (a): A detected stick event; an upper threshold of 0.04 rad/s (dash-dot, green) is used to define stick events.}
    \label{fig:samplespeed}
\end{figure*}

To eliminate basal friction in one set of experiments, particles are coated on the top surface with a superhydrophobic layer and float on de-ionized water, as  shown in the inset of Fig.~\ref{fig:schematic}(b). Static friction coefficients $\mu$ for dry experiments with basal friction were measured through tilt tests with particles in contact with other particles or with a glass base. For floated particles, $\mu$ was measured by recording the force required to slide disks past one another when resting on top of each other under gravity with edge-to-edge contacts.  The contact line was under water to account for fluid lubrication at the contacts. 

In our experiments, $\phi$ is varied by changing the number of particles in the annular channel while approximately fixing the small to large particle number ratio. With increasing $\phi$, we observe stick events that reach higher angular displacements. Thus the upper limit $\phi_\text{max}$ for each experiment with basal friction ($\phi_\text{max}=0.772$ with high interparticle friction, $\phi_\text{max}=0.796$ with low interparticle friction) is set by the maximum angular displacement by which the spring can be compressed in the apparatus ($180^{\circ}$) as well as an increasing probability for particles to buckle out of plane as the average force in the medium increases. When there is no basal friction, we do not observe high angular displacement stick events at all; the upper limit ($\phi_\text{max}=0.785$) is instead set by the increasing probability for particles to buckle out of plane, which is more likely to occur in the floating system even at low forces because particles at the fluid-air interface do not have a fixed substrate to prevent buckling in one direction. The lower limit $\phi_\text{min}$ when basal friction is present is the packing fraction below which the intruder forms a narrow open channel through which to glide and does not interact with the granular medium in the steady state. We find $\phi_\text{min}=0.650$ and $\phi_\text{min}=0.677$ for high and low interparticle friction, respectively. Without basal friction, $\phi_\text{min}=0.750$ corresponds to a steady state in which particles simply glide around the intruder with few force-bearing interactions\footnote{Without basal friction, the particles experience mutual capillary attractions because they are floating on water. Thus, a channel never forms at very low $\phi$ as happens in the cases with basal friction.}. We note that the ranges of $\phi$ attainable are either below or on the low end of the range of jamming packing fractions between random loose packing $\phi_{\text{RLP}} \approx 0.77$ \cite{RLP3Donoda,RLP2D3Dsilbert} and 
random close packing $\phi_{\text{RCP}} \approx 0.84$ \cite{RCPtorquato,RCP3Dscott}, in agreement with other intruder experiments with frictional particles in a similar geometry \cite{slowdraggeng}. In other \textit{linear} intruder experiments \cite{dragforcecavityformationintruderkolb,intrudervibrationgdauchot} (1) there was no limiting displacement of a driving spring and (2) the system size was significantly larger than here (the intruder was a smaller perturbation), so values of $\phi$ significantly closer to $\phi_{\text{RCP}}$ were accessible.

\section{Results \label{sec:results}}

\subsection{Qualitative Observations}

Figure~\ref{fig:samplespeed} displays sample time series of the velocity of the intruder and the force of the grains on the intruder in the experiments with and without basal friction at comparable interparticle friction. Plots are shown for $\phi_{\text{min}}$ and $\phi_{\text{max}}$ for the two sets of experiments. Both the force and velocity are plotted with respect to cumulative drive angle, proportional to time by $\dot{\theta}_{d}$. 

\textit{With basal friction}: At low $\phi$, the intruder spends most of the time moving at velocities near $\dot{\theta}_{d}$ with fluctuations but occasionally gets stuck, having a velocity near zero (Fig.~\ref{fig:samplespeed}(a)). The small fluctuations about the drive velocity are caused by the intruder pushing loose grains or clusters of grains out of its way, the transient formation of weak force chains in the granular medium that break before stopping the intruder, or driving fluctuations ($\pm 0.1 \text{ rad/s}$) due to friction in the central post. The force also exhibits fluctuations on the order of $\pm 0.05 \text{ N}$ when the intruder is moving near the drive velocity and a noise level of $10^{-3} \text{ N}$ when the intruder is stuck (nearly stationary). As $\phi$ increases, stick events occur more frequently until beyond a certain range of $\phi$ the intruder exhibits pure \textit{stick-slip} dynamics dominated by rapid slip events followed by relatively long stick events (Figs.~\ref{fig:sampletimeseriesplots}(a) and~\ref{fig:samplespeed}(b)). During stick events, the force on the intruder increases approximately linearly with time as the torque spring is compressed at a constant rate. During slips at high $\phi$, the force fluctuates rapidly and the peak velocity is generally set by the amount of compression in the spring upon release --- greater compression in a stick event leads to a more rapid slip event. The low-$\phi$ dynamics observed in Fig. \ref{fig:samplespeed}(a) will be called \textit{intermittent flow} (or clogging-like) dynamics. The high-$\phi$ dynamics observed in Fig. \ref{fig:samplespeed}(b) will be called \textit{stick-slip} dynamics. 

\textit{Without basal friction}: For \textit{all} $\phi$ studied in this experiment, the system exhibits intermittent flow with the intruder moving near the drive velocity most of the time (Figs.~\ref{fig:sampletimeseriesplots}(b) and~\ref{fig:samplespeed}(c,d)). 
Note that the vertical scales for both velocity and force in Fig.~\ref{fig:samplespeed} are smaller in (c,d) compared to those in (a,b).

\subsection{Intruder Velocity and Force Distributions \label{velforcedists}}

\begin{figure}
    \centering
    \includegraphics[width=1\linewidth]{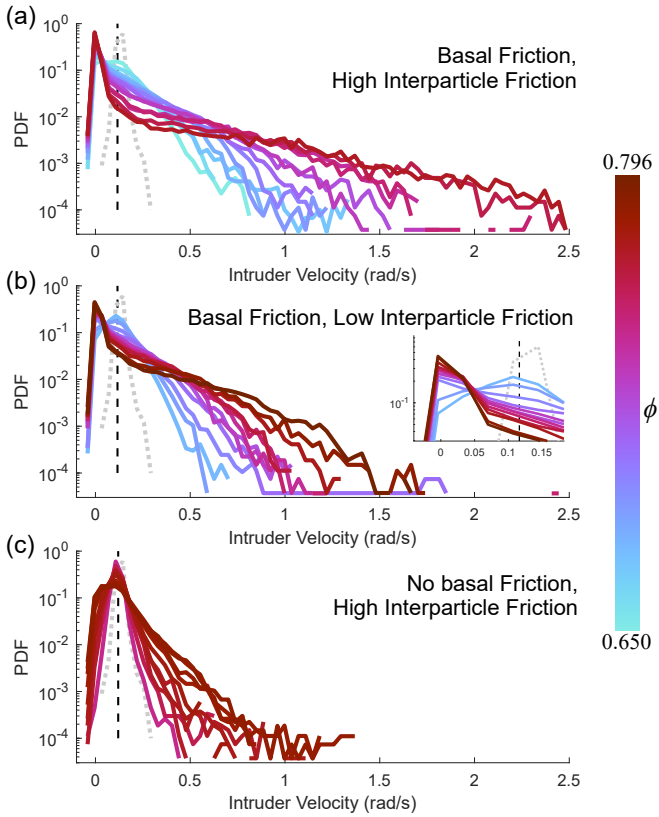}
    \caption{(Color online.) Velocity distributions (bin count normalized to total number of data points) with varying packing fraction, $\phi$. Vertical black dashed lines indicate $\dot{\theta}_{d}$. The dotted (gray) distribution is noise measured by driving the intruder around the annulus without any particles. (a) Basal friction, high interparticle friction. (b) Basal friction, low interparticle friction. (c) No basal friction, high interparticle friction. Inset of (b): Zoom in of plot to highlight the increase in the peak height at 0 rad/s with increasing $\phi$; this trend is observed for all three experiments.}
    \label{fig:speeddist}
\end{figure}

Figure~\ref{fig:speeddist} shows the probability distributions (PDF) of intruder velocities for several values of $\phi$. 

\textit{With basal friction} (Fig.~\ref{fig:speeddist}(a,b)): At low $\phi$, the distribution has a significant peak at $\dot{\theta}_{d}$. As $\phi$ increases, the height of this peak diminishes as a peak at zero velocity develops, showing that the intruder is stuck for larger fractions of the total time as $\phi$ increases. The distribution also broadens, as faster slip velocities are achieved after stick events with greater spring deflection. The velocity distributions peaked at $\dot{\theta}_{d}$ correspond to the intermittent flow dynamics, whereas the distributions where the dominant peak is at zero are consistent with fully developed stick-slip dynamics. The data suggest a smooth crossover between the two regimes rather than a sharp transition. For the lower interparticle friction case (Fig.~\ref{fig:speeddist}(b)), these trends are qualitatively the same, with the crossover occurring at higher values of $\phi$ and a substantial decrease in the maximum observed velocities for a given $\phi$.

\textit{Without basal friction} (Fig.~\ref{fig:speeddist}(c)): The distribution has a significant contribution at $\dot{\theta}_{d}$ for all accessible $\phi$. With increasing $\phi$, the distribution broadens as longer sticks and associated faster slips occur, and weight is shifted to zero, indicating more and/or longer stick events. However, even at the largest packing fraction $\phi_{\text{max}}$ attainable in the experiment, zero velocity does not dominate, indicating that a fully developed stick-slip dynamics has not been achieved.

In all cases, some negative velocities are observed in the signal. In the intermittent flow regime, the spring is often able to completely decompress in a slip event, causing the driving mechanism wedges shown in Fig.~\ref{fig:torquespring} to collide and the intruder to rebound. In the stick-slip regime with basal friction, the granular medium sometimes does push the intruder backwards at the end of a slip event.

\begin{figure}
    \centering
    \includegraphics[width=1\linewidth]{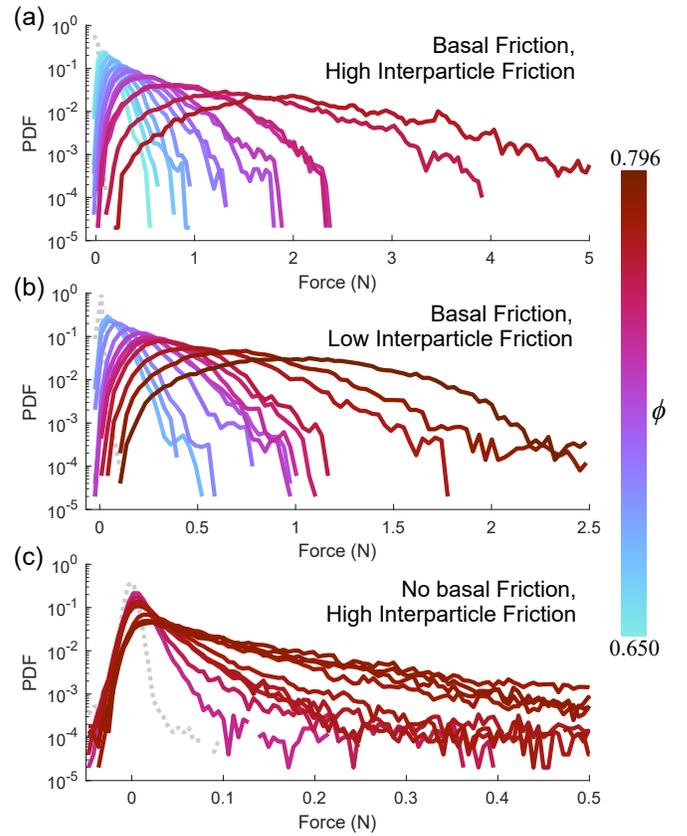}
    \caption{(Color online.) Force distributions with varying packing fraction, $\phi$. The dotted (gray) distribution is noise measured by driving the intruder around the annulus without any particles. (a) Basal friction, high interparticle friction. (b) Basal friction, low interparticle friction. (c) No basal friction, high interparticle friction.}
    \label{fig:forcedist}
\end{figure}

Figure~\ref{fig:forcedist} shows the distributions of forces. Note the differences in the horizontal scales of the three panels. Both with basal friction (a,b) and without basal friction (c), the maximum of the PDF shifts to higher forces as $\phi$ increases. The width of the distribution increases with $\phi$ in all three cases but decreases substantially with lower interparticle friction and even more dramatically when basal friction is eliminated. This suggests a decrease in the number and duration of stick events in the two cases. Figure~\ref{fig:statisticsplots}(a) summarizes the information from the force distributions, showing the average force and distribution width for all three experimental conditions and packing fractions.

The observations from the velocity and force PDFs suggest that at higher $\phi$, where there is less area available for grains to flow behind the intruder \cite{dragforcecavityformationintruderkolb}, grains are more likely to form mechanically stable structures that cause the intruder to get stuck \cite{statphysgranularTighe}.  In the fully developed stick-slip regime, the fact that increasing $\phi$ leads to longer stick durations implies the formation of force chains with greater mechanical stability,  consistent with previous studies in which the critical force that an intruder needed to break stable granular structures increased with $\phi$ \cite{probeintruderreichhardt,slowdraggeng,intrudervibrationgdauchot}.

\subsection{Analysis of Stick Events}\label{sec:stick}

In Fig.~\ref{fig:samplespeed}, detected stick events are shown in black. The inset in Fig.~\ref{fig:samplespeed}(a) shows the threshold of 0.04 rad/s used to define stick events; we ignore any microslip events that may occur below this threshold during a stick event. Detected events that are shorter than 0.40 seconds (or a cumulative drive angle of $\sim 0.04$ rad, or 20 data points) are also disregarded. For each detected event, $\theta(t)$ during that time interval is fitted with a line whose slope is the representative average \textit{stick creep velocity} of that stick event. The duration of each event is also recorded, and by multiplying this duration by $\dot{\theta}_{d}$ we obtain an approximate \textit{amplitude} of the event (the total spring deflection during the event) in radians. We checked the return map (not shown) for consecutive stick event amplitudes and found no indication of periodic dynamics even in the stick-slip regime. 

\begin{figure}
    \centering
    \includegraphics[width=\linewidth]{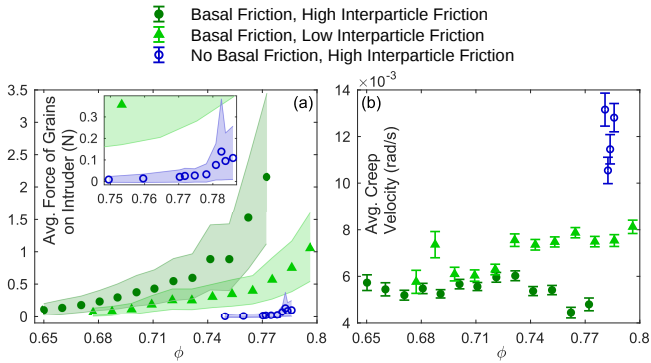}
    \caption{(Color online.) (a) The average force of the force time series as a function of packing fraction, $\phi$, for each experimental condition. Each shaded region around a point ranges from the lower 10\% cutoff to the upper 90\% cutoff of that point's force distribution. Inset: Zoom in to highlight the case with no basal friction. (b) The average creep velocity of the intruder during detected stick events as a function of $\phi$ (the average speeds associated with the thick black lines in Fig.~\ref{fig:samplespeed}).}
    \label{fig:statisticsplots}
\end{figure}

Figure~\ref{fig:statisticsplots}(b) shows the average of all stick creep velocities for each $\phi$ and frictional condition. Here, we only display data points for sets containing at least 20 stick events, a criterion which excludes low $\phi$ experiments. With basal friction and high interparticle friction, the creep velocity is lowest; with lower interparticle friction the creep velocity is a little higher at high $\phi$. Without basal friction, the creep velocity is roughly twice that of the experiments with basal friction. The fact that the creep velocity measurably changes with frictional properties suggests that the granular medium rearranges on a small scale to accommodate for the driven intruder, possibly through many separate microslips or a more continuous chain of rearrangement events. 

The intermittent flow and stick-slip regimes (see Fig.~\ref{fig:introdiagram}) can be distinguished by considering the statistics of the waiting time, $\tau_w$, between stick events, where $\tau_w$ is defined to be the time from the end of one stick event to the beginning of the next (see Fig.~\ref{fig:samplespeed}), including the duration of the associated slip between the two stick events. If the intermittent flow dynamics are analogous to clogging, then stick events will follow a Poissonian distribution and the waiting time distribution is expected to be exponential consistent with the distribution of avalanche sizes in clogging \cite{clogconfigsdurian,zuriguel2014clogging,patterson2017clogging,zuriguel2005jamming}. By contrast, in the stick-slip regime, $\tau_w$ is expected to be a small value with a small distribution width set only by the duration of slip events, the longest of which are on the order of the inertial timescale of the cantilever. 

\begin{figure*}
    \centering
    \includegraphics[width=1.0\linewidth]{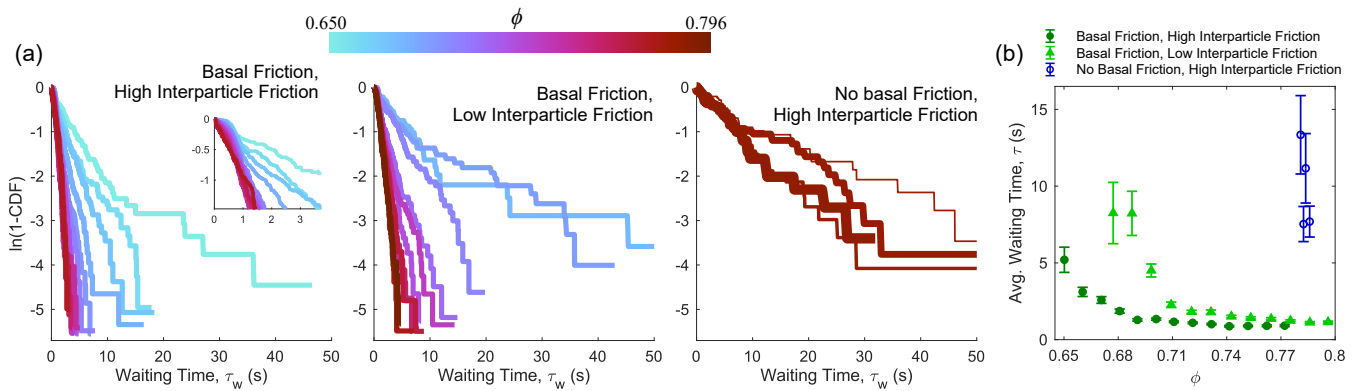}
    \caption{(Color online.) (a) The complementary cumulative distribution function (CDF) of the distribution of waiting times, $\tau_w$, between stick events for all three experiments. The inset is a zoomed in plot to highlight the trend of the complementary CDFs at low $\tau_w$. \textit{Without basal friction}: increasing thickness here denotes increasing packing fraction $\phi$: $0.781, 0.783, 0.784, 0.786$. (b) The average waiting time $\tau \equiv \langle \tau_w \rangle$ with error bars given by the standard error of the average.}
    \label{fig:taudist}
\end{figure*}

The cumulative distribution functions (CDF) of $\tau_w$ for all three experiments at varying $\phi$ are shown in Fig.~\ref{fig:taudist}(a).
The distributions do not seem to be exponential at lower $\phi$, as one would expect if the system were analogous to clogging systems. Below $\tau_w \approx 15 \text{ s}$ for the lowest $\phi$'s, the distributions with basal friction do appear to be exponential, but the waiting times above 15 s are more numerous than expected for the exponential trend. Long waiting times usually occur at such low $\phi$ when the intruder nearly forms a free channel in the granular medium that extends over a significant fraction of its path around the annulus. In such cases, the intruder can  move freely or simply push along one particle without interacting with the medium as a whole until it encounters a constricted portion of the channel and resumes the intermittent flow with shorter waiting times.

Without basal friction, all distributions are statistically identical for the $\phi$'s at which we were able to obtain a significant number of stick events.  The distributions are roughly exponential, but the numbers of events are too low to support firm conclusions. In any case, the trend of faster decay of the waiting time probability with increasing $\phi$ is apparent for both experiments with basal friction. We utilize this decrease in the \textit{average waiting time} for a given experiment $\tau \equiv \langle \tau_w \rangle$ to distinguish the dynamical regimes of stick-slip and intermittent flow (Fig.~\ref{fig:taudist}(b)) --- when the average waiting time becomes comparable to the inertial timescale of the cantilever, the distribution decays sharply and $\tau$ is on the order of $1\,$s, indicating stick-slip dynamics. (We do not attempt here to propose a quantitative form for the decay or to assign to it a precise value of the packing fraction.) A distribution with waiting times significantly longer than the inertial timescale indicates intermittent flow. We schematically summarize the results of Fig.~\ref{fig:taudist}(b) in Fig.~\ref{fig:introdiagram}, with the crossovers between stick-slip and intermittent flow for the cases with basal friction occurring at $\phi \approx 0.70$ for high interparticle friction and $\phi \approx 0.73$ for low interparticle friction. The experiment without basal friction does not exhibit stick-slip dynamics for the $\phi$ accessible in this experiment, though we suspect that a crossover would occur at $\phi$ nearer to $\phi_{\text{RCP}}$ if buckling out of plane could be avoided.

\section{\label{sec:discussion} Discussion and Conclusions}

We have collected data on the dynamics of a grain-scale intruder pushed through quasi-2D granular media using a spring of finite stiffness and varying interparticle friction and basal friction of the grains.  The annular geometry of the system enables us to study statistically steady states for a range of packing fractions and several values of the static friction coefficients.

We find that the dynamics are strongly affected by whether basal friction is present or not. With basal friction and increasing $\phi$, the intruder experiences a smooth crossover in dynamics from intermittent flow to stick-slip as revealed by smooth crossovers in both the average force (Fig.~\ref{fig:statisticsplots}(a)) and average waiting time between stick events (Fig.~\ref{fig:taudist}(b)). In the intermittent flow regime, the intruder spends most of the time moving through the medium with small fluctuations in velocity due to collisions and short-lived, weak force chains, but occasionally gets stuck for extended periods. The motion resumes when the force exerted on the medium by the intruder increases beyond a yield threshold that differs for each stick event. In the stick-slip regime, the intruder motion is dominated by quick slips following stick events.  These slips feature rapidly varying interactions between the intruder and grains. Interparticle friction $\mu$ controls the range of $\phi$ over which the crossover from intermittent flow to stick-slip occurs, with the crossover occurring over a lower range of $\phi$ for larger $\mu$.  When basal friction is eliminated, the intermittent flow regime extends to all packing fractions covered by our experiments, including those where stick-slip was observed in the presence of basal friction. 

\begin{figure}
    \centering
    \includegraphics[width=0.6\linewidth]{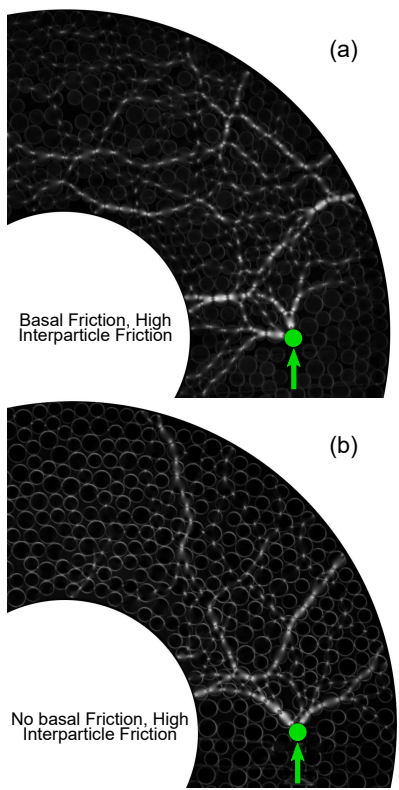}
    \caption{(Color online.) Sample photoelastic images of representative stick events at $\phi = 0.762$ (a) and $\phi = 0.780$ (b) and a measured force of $\sim 0.8 \text{ N}$ for the experiments with basal friction and without basal friction at high interparticle friction. The intruder is marked with a circle, and the arrow pointing to the intruder indicates the direction of force applied to the intruder by the torque spring.}
    \label{fig:samplephotoelastic}
\end{figure}

Our results raise several questions for future studies. First, it would be interesting to see how changes in the drive velocity or the spring constant modify the dynamics. In a variety of granular stick-slip experiments and simulations with applied boundary shear or large sliders, these parameters control the macroscopic dynamical regime \cite{analyticavalanchesdahmen,granularfaultdanielshayman,stickslipcracklingagheal,stickslipalbert,stickslipnasuno}. One may ask, for example, whether observed effects such as periodic stick-slip dynamics can occur when the interaction of the drive with the medium is confined to the single grain scale.

Second, one may ask how the intruder dynamics at low $\phi$ is related to the clogging phenomena observed in granular flows through small apertures. The intruder acts as a boundary between the inner and outer ring, creating two effective apertures.  Our system is thus related to a highly coupled two-aperture clogging system \cite{multipleorificesKunte2014}, which is less well understood than the single-aperture case \cite{clogto,clogconfigsdurian,zuriguel2005jamming}. Our system sticks (or clogs) more often than expected for a single aperture of a size equivalent to the width of one of our two "effective apertures," which are each approximately 6 large particle diameters across. A simple, and likely naive, argument suggests that the clogging probability of our system, $p_\textrm{annulus}$, should be connected with the probability of clogging at least one of two apertures; $p_\textrm{annulus}=2p-p^2$, where $p$ is the probability of clogging a single aperture and the clogging events for the apertures are assumed to be equal and uncorrelated.  This leads us to expect $p_\textrm{annulus} > p$, in agreement with the observations of stick event frequencies in the intermittent flow regime and especially with the observation of any stick events in the absence of basal friction.  To test the relevance of clogging concepts to our stick events it will be crucial to gather data for different channel widths or grain sizes. 

Third, we have seen that the creep velocities of the intruder during nominal stick events depend on the interparticle and basal frictional properties of the granular material, as shown in Fig.~\ref{fig:statisticsplots}(b). Why does lower interparticle friction or reduced basal friction increase the average creep velocity?  Acquiring data on the microscopic motions of grains during creep would help answer this question by distinguishing between the occurrence of multiple, discrete microslip events and continuous bulk rearrangements.  Measurements of force chain networks may also reveal differences between the sticking configurations achieved in systems with different frictional properties. 

Though this paper did not emphasize it, our experimental system allows us to gather photoelastic data that reveals internal stress structures in the granular medium during intruder sticks and slips. For example, Fig.~\ref{fig:samplephotoelastic} shows two representative experimental images, one from a run with basal friction and the other from a run with no basal friction. The measured force on the intruder was the same for the two images. In the case with basal friction, some force chains extend behind the intruder (the intruder is being driven in the direction of the arrow by the torque spring), and this is typical. In contrast, without basal friction, no force chains that extend behind the intruder have been observed during any stable stick events. Future studies, which are beyond the scope of this paper, aim to provide quantitative statistical analyses of these and other features of stress states and relaxation dynamics within the granular medium.

Finally, while we have emphasized the distinctions between the stick-slip and clogging regimes, the smoothness of the crossover between the two suggests a possible relation between them.  Future studies of the spatial distributions of force chains during clogging and sticking events may reveal links between these phenomena and the shear jammed or fragile states on the yield curve of the phase diagram recently observed in a Couette experiment driven by globally applied shear~\cite{YiqiuPRL}
\vspace{11pt}
\begin{acknowledgments}
This work was supported by the US Army Research Office through grant W911NF1810184 and by the Keck Foundation.  L. A. P. and C. M. C. acknowledge support by Universidad Tecnol\'ogica Nacional through grants PID-MAUTNLP0004415 and PID-MAIFIBA0004434TC and CONICET through grant RES-1225-17. C. M. C. also thanks the Norma Hoermann Foundation for partial funding for his visit to NJ. L. K. was supported in part by NSF Grant No. 1521717. 
\end{acknowledgments}

\end{document}